\RequirePackage{lineno}
\documentclass[twocolumn,showpacs,preprintnumbers,amsmath,amssymb,prl,superscriptaddress]{revtex4}
\usepackage{graphicx}
\usepackage{dcolumn}
\usepackage{bm}
\usepackage{hyperref}
\modulolinenumbers[1]
\newenvironment{color}[3]
{
}{
}

\newcommand{\grey}[1]     {}

\newcommand{\pT}{$p_{\rm{T}}$}

\newcommand{\muB}{$\mu_{\rm B}$}
\newcommand{\sNN}{$\sqrt {{s_{\rm NN}}}$}
\newcommand{\Npart}{$\langle N_{{\rm part}}  \rangle$}
\newcommand{\MV}{$\sigma^{2}/M$}
\newcommand{\Ss}{$S\sigma$}
\newcommand{\KV}{$\kappa\sigma^{2}$}

\begin{document}
\title{ Beam energy dependence of moments 
of the net-charge multiplicity distributions in Au+Au collisions at RHIC }
\affiliation{AGH University of Science and Technology, Cracow, Poland}
\affiliation{Argonne National Laboratory, Argonne, Illinois 60439, USA}
\affiliation{University of Birmingham, Birmingham, United Kingdom}
\affiliation{Brookhaven National Laboratory, Upton, New York 11973, USA}
\affiliation{University of California, Berkeley, California 94720, USA}
\affiliation{University of California, Davis, California 95616, USA}
\affiliation{University of California, Los Angeles, California 90095, USA}
\affiliation{Universidade Estadual de Campinas, Sao Paulo, Brazil}
\affiliation{Central China Normal University (HZNU), Wuhan 430079, China}
\affiliation{University of Illinois at Chicago, Chicago, Illinois 60607, USA}
\affiliation{Cracow University of Technology, Cracow, Poland}
\affiliation{Creighton University, Omaha, Nebraska 68178, USA}
\affiliation{Czech Technical University in Prague, FNSPE, Prague, 115 19, Czech Republic}
\affiliation{Nuclear Physics Institute AS CR, 250 68 \v{R}e\v{z}/Prague, Czech Republic}
\affiliation{Frankfurt Institute for Advanced Studies FIAS, Germany}
\affiliation{Institute of Physics, Bhubaneswar 751005, India}
\affiliation{Indian Institute of Technology, Mumbai, India}
\affiliation{Indiana University, Bloomington, Indiana 47408, USA}
\affiliation{Alikhanov Institute for Theoretical and Experimental Physics, Moscow, Russia}
\affiliation{University of Jammu, Jammu 180001, India}
\affiliation{Joint Institute for Nuclear Research, Dubna, 141 980, Russia}
\affiliation{Kent State University, Kent, Ohio 44242, USA}
\affiliation{University of Kentucky, Lexington, Kentucky, 40506-0055, USA}
\affiliation{Korea Institute of Science and Technology Information, Daejeon, Korea}
\affiliation{Institute of Modern Physics, Lanzhou, China}
\affiliation{Lawrence Berkeley National Laboratory, Berkeley, California 94720, USA}
\affiliation{Massachusetts Institute of Technology, Cambridge, MA 02139-4307, USA}
\affiliation{Max-Planck-Institut f\"ur Physik, Munich, Germany}
\affiliation{Michigan State University, East Lansing, Michigan 48824, USA}
\affiliation{Moscow Engineering Physics Institute, Moscow Russia}
\affiliation{National Institute of Science Education and Research, Bhubaneswar 751005, India}
\affiliation{Ohio State University, Columbus, Ohio 43210, USA}
\affiliation{Old Dominion University, Norfolk, VA, 23529, USA}
\affiliation{Institute of Nuclear Physics PAN, Cracow, Poland}
\affiliation{Panjab University, Chandigarh 160014, India}
\affiliation{Pennsylvania State University, University Park, Pennsylvania 16802, USA}
\affiliation{Institute of High Energy Physics, Protvino, Russia}
\affiliation{Purdue University, West Lafayette, Indiana 47907, USA}
\affiliation{Pusan National University, Pusan, Republic of Korea}
\affiliation{University of Rajasthan, Jaipur 302004, India}
\affiliation{Rice University, Houston, Texas 77251, USA}
\affiliation{Universidade de Sao Paulo, Sao Paulo, Brazil}
\affiliation{University of Science \& Technology of China, Hefei 230026, China}
\affiliation{Shandong University, Jinan, Shandong 250100, China}
\affiliation{Shanghai Institute of Applied Physics, Shanghai 201800, China}
\affiliation{SUBATECH, Nantes, France}
\affiliation{Temple University, Philadelphia, Pennsylvania, 19122, USA}
\affiliation{Texas A\&M University, College Station, Texas 77843, USA}
\affiliation{University of Texas, Austin, Texas 78712, USA}
\affiliation{University of Houston, Houston, TX, 77204, USA}
\affiliation{Tsinghua University, Beijing 100084, China}
\affiliation{United States Naval Academy, Annapolis, MD 21402, USA}
\affiliation{Valparaiso University, Valparaiso, Indiana 46383, USA}
\affiliation{Variable Energy Cyclotron Centre, Kolkata 700064, India}
\affiliation{Warsaw University of Technology, Warsaw, Poland}
\affiliation{University of Washington, Seattle, Washington 98195, USA}
\affiliation{Yale University, New Haven, Connecticut 06520, USA}
\affiliation{University of Zagreb, Zagreb, HR-10002, Croatia}

\author{L.~Adamczyk}\affiliation{AGH University of Science and Technology, Cracow, Poland}
\author{J.~K.~Adkins}\affiliation{University of Kentucky, Lexington, Kentucky, 40506-0055, USA}
\author{G.~Agakishiev}\affiliation{Joint Institute for Nuclear Research, Dubna, 141 980, Russia}
\author{M.~M.~Aggarwal}\affiliation{Panjab University, Chandigarh 160014, India}
\author{Z.~Ahammed}\affiliation{Variable Energy Cyclotron Centre, Kolkata 700064, India}
\author{I.~Alekseev}\affiliation{Alikhanov Institute for Theoretical and Experimental Physics, Moscow, Russia}
\author{J.~Alford}\affiliation{Kent State University, Kent, Ohio 44242, USA}
\author{C.~D.~Anson}\affiliation{Ohio State University, Columbus, Ohio 43210, USA}
\author{A.~Aparin}\affiliation{Joint Institute for Nuclear Research, Dubna, 141 980, Russia}
\author{D.~Arkhipkin}\affiliation{Brookhaven National Laboratory, Upton, New York 11973, USA}
% ...

\bigskip
\collaboration{STAR Collaboration}\noaffiliation
\bigskip
\date{ \today }

\begin{abstract}

We report the first measurements of the moments \textemdash~mean ($M$), variance
($\sigma^{2}$), skewness ($S$) and kurtosis ($\kappa$) \textemdash~of
the net-charge multiplicity distributions at mid-rapidity in Au+Au collisions at seven
energies, ranging from \sNN~=~7.7 to 200~GeV,
as a part of the Beam Energy Scan program at RHIC. 
The moments are related to the thermodynamic susceptibilities of net-charge, 
and are sensitive to the location of the QCD critical point.
We compare the products of the moments,
\MV, \Ss~and \KV~with the expectations from
Poisson and negative binomial distributions (NBD).
The \Ss~values deviate from Poisson and are close to NBD baseline, while
the \KV~values tend to lie between the two.
Within the present uncertainties, our data do not show
non-monotonic behavior as a function of collision energy. 
These measurements provide a valuable tool to extract the
freeze-out parameters in heavy-ion collisions by comparing with
theoretical models.

\end{abstract}

\maketitle

The major goals of the physics program at Brookhaven National Laboratory's Relativistic
Heavy-Ion Collider (RHIC) are the search and study  
of a new form of matter known as the Quark-Gluon Plasma
(QGP)~\cite{RHICQGP} and the 
mapping of the Quantum Chromodynamics (QCD) phase diagram 
in terms of temperature ($T$) and baryon chemical potential (\muB). 
Lattice QCD calculations indicate that at vanishing \muB, the
transition from the QGP to a hadron gas is a smooth 
crossover~\cite{aoki,ejiri,bowman,stephanov,fodor,gavai,cheng}, 
while at large \muB, the phase transition is
of first order~\cite{ejiri,herold}. 
Therefore, a critical point in the QCD phase diagram is expected at
finite \muB, where the first order transition ends. The location of the critical point has
been predicted to be accessible at
RHIC~\cite{steph,rajagopal,STAR_BES}, where the Beam Energy Scan program has been
ongoing since 2010. The aim of this program is to 
map the QCD phase by varying the center-of-mass energy of the colliding ions,
thereby scanning a large window in \muB~and $T$. 

One of the characteristic signatures of the QCD critical point is
the non-monotonic behavior in the fluctuations of globally conserved quantities,
such as net-baryon, net-charge, 
and net-strangeness number as a function of beam 
energy~\cite{athanasiou,stephanov,steph,rajagopal,STAR_BES,STAR2,ejiri,fodor,gavai,cheng,skokov,PBM,asakawa,kitazawa,bzdak}. 
The event-by-event distributions of the conserved quantities within a
limited acceptance are characterized by the moments, 
such as the mean ($M$), the standard deviation ($\sigma$), the
skewness ($S$) which represents the asymmetry of the distribution,
and the kurtosis
($\kappa$) which gives the degree to which the distribution is peaked
relative to the normal distribution.
These moments are
related to the corresponding higher-order thermodynamic susceptibilities and to the
correlation length of the 
system~\cite{stephanov2009,athanasiou}. 
At the critical point, thermodynamic susceptibilities and the
correlation length of the system are expected to diverge for large
samples in equilibrium. But in
reality, the phenomenon of critical slowing down in the vicinity of the critical point drives the system away
from thermodynamic equilibrium, so the correlation length 
reaches a maximum value of around $1.5-3$~fm~\cite{berdnikov,athanasiou}. 
Assuming that the signal
at freeze-out survives dissipation during the evolution of the fireball from
the hadronization stage~\cite{evolution}, the higher moments can be
used as one of the preferred tools for locating the critical point.

When relating the susceptibilities to the moments, a
volume term appears, making it difficult to compare
different systems and collision centralities. 
The products of the moments, such as \MV, \Ss~and \KV, are
constructed in order to cancel the volume term.
Lattice QCD calculations have 
shown that these products go through rapid change near the
critical point~\cite{ejiri,fodor,gavai,cheng}.
In addition, the products of the moments of the experimental data 
can be effectively used to determine the freeze-out points on the QCD
phase diagram
by comparing directly with first-principle lattice QCD calculations~\cite{freezeout_karsch}.
The net-charge multiplicity distributions are appropriate for 
all these studies as they directly probe a conserved
quantum number~\cite{bzdak,skokov,kitazawa}. 
Combining these results with the moments of 
net-proton multiplicity distributions~\cite{star_net_proton}, we may be
able to extract the freeze-out parameters and probe the critical point.

In this Letter, we report the first measurements of the moments of the
net-charge multiplicity distributions in Au+Au collisions at
$\sqrt{s_{NN}}$ = 7.7, 11.5, 19.6, 27, 39, 62.4 and 200~GeV,
corresponding to \muB~from 410 to 20 MeV~\cite{Cleymans}.  

The data were taken by the Solenoid Tracker at RHIC (STAR) experiment
in 2010 and 2011, as part of the Beam Energy Scan program at 
RHIC~\cite{STAR2,STAR_BES,daniel,nihar_thesis}.
With large uniform acceptance and excellent particle identification capabilities, STAR provides an
ideal environment for studying event-by-event distributions of charged
particles. The Time Projection Chamber (TPC)~\cite{TPC} is the main
tracking detector used to identify charged
particles and obtain net-charge (difference between the number of positive and
negative charged particles) on an event-by-event basis.
Combination of signals from the Zero Degree Calorimeters~\cite{ZDC}, Vertex
Position Detectors~\cite{VPD} and Beam-Beam Counters~\cite{BBC} are used as the
minimum-bias trigger.
The data analysis has been carried out for collisions occurring within $\pm$30
cm of the TPC center in the beam direction. 
Interactions of the beam with the beam pipe are rejected by selecting
events with a radial vertex position in the transverse plane of less
than 2~cm. The charged tracks are selected
with more than 20 space points in the TPC 
out of 45, 
a distance of closest approach (DCA) to the primary vertex of less than 1~cm and
number of hit points used to calculate the specific energy loss
greater than 10. 
The spallation protons, produced due to beam-pipe
interactions, affect the charged particle measurement. These are
suppressed by removing protons with transverse momentum (\pT) less
than 400 MeV/{\it c}. To be consistent,
anti-protons are also removed within this \pT~range.
The centrality of the collision is determined by using
the total number of charged particles within a pseudorapidity ($\eta$)
window of
$0.5<|\eta|< 1.0$, chosen to be beyond the analysis window of the
net-charge distributions. 
The centrality is represented by the average number of participating nucleons
(\Npart) as well as percentage of total cross section,
obtained by the Monte Carlo (MC) Glauber simulation~\cite{Michael_Glauber}.
The total number of events analyzed 
are (in millions): 1.4, 2.4, 15.5, 24, 56, 32 and 75 for 
\sNN~= 7.7, 11.5, 19.6, 27, 39, 62.4 and 200~GeV, respectively.
   
The measured positive ($N_{+}$) and negative ($N_{-}$)
charged particle multiplicities within  $|\eta|<0.5$ and 
$0.2 < p_{\rm{T}} < 2.0$~GeV/{\it{c}} 
(after removing protons and anti-protons with $ p_{\rm{T}} <
400$~MeV/{\it c}) are used to calculate net-charge ($ N_{+} - N_{-}$) in each event.
The net-charge distributions are obtained for different centrality classes.
The finite centrality bin width may cause volume variations within a given
centrality class and may introduce additional fluctuations. 
The moments and moments products are calculated at every integer value
of the centrality variable. The values shown in the figures are weighted
averages in 5\% or 10\% wide centrality bins, where the weights are
the number of events at each value of the centrality variable normalized
to unity within each such centrality bin. Such weighted averages effectively
remove the dependence of the results on the width of the centrality
bin~\cite{CBW,nihar}. 
The correction factor on the higher moments for choosing 5\% or 10\%
wide centrality bins compared to narrower (1\%) bins is about 2\%
or less.

Finite reconstruction efficiencies of the charged particles affect the
measured moments.
The efficiency for each centrality and collision
energy is obtained by using the embedding technique~\cite{embed}.
The average efficiencies 
vary within 63\%$-$66\% and 70\%$-$73\% for most central (0-5\% bin) and
peripheral (70-80\% bin) events, respectively, for all collision energies. 
The corrections to the moments 
are based on binomial probability distributions of
efficiency~\cite{bzdak}. For \KV, the efficiency correction factors for all energies and
centralities are consistent with unity, whereas for \Ss, these factors
vary from 1.4 to 1.0 from peripheral to 
central collisions for all energies.

The statistical errors of the moments and their products have been
calculated using the Delta theorem approach~\cite{Luo} and Bootstrap
method~\cite{bootstrap} for efficiency-uncorrected and corrected
results, respectively. 
The statistical uncertainties in the corrected results increase
compared with the uncorrected ones
because the efficiency corrections involve higher-order cumulants.
The systematic uncertainties are obtained by
varying the track selection criteria of the charged particles,
such as the number of fit points, DCA, 
and the number of hit points used to calculate ionization energy loss
({\it dE/dx})  in the TPC. The final systematic errors were estimated by including
an additional 5\% uncertainty in the reconstruction efficiency.

\begin{figure}[tbp]
\centering
\includegraphics[width=0.52\textwidth]{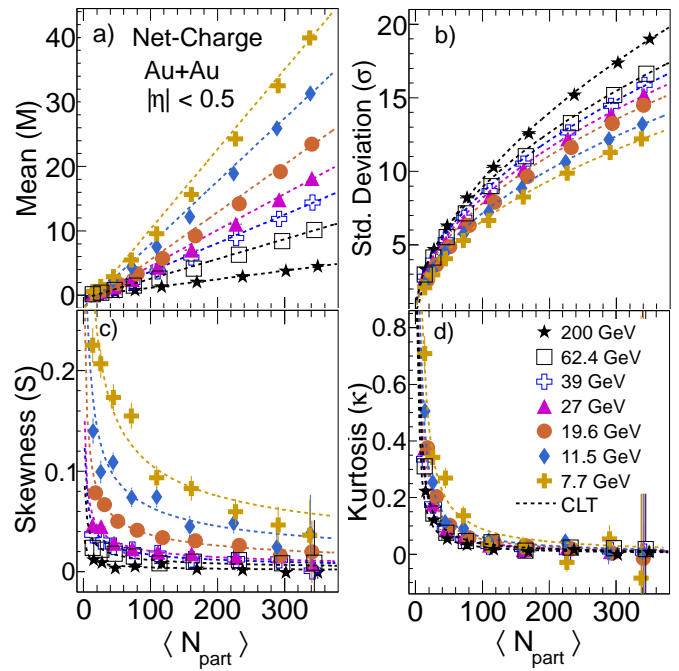} 
\caption{(Color online) The efficiency and centrality bin width
  corrected (a) mean, (b) standard
deviation, (c) skewness and (d) kurtosis of the net-charge multiplicity
distributions as a function of 
number of participating nucleons (\Npart)~for Au+Au collisions.  
The dotted lines represent
calculations from the central limit theorem.
The error bars are statistical and systematic errors are within the symbol sizes.}
\label{fig1}
\end{figure}

In Fig.~\ref{fig1}, the efficiency and centrality bin width 
corrected moments of the net-charge
distributions are plotted as a function of \Npart~for Au+Au collisions
at seven colliding energies.
The statistical errors dominate in most cases and the systematic
errors are within the symbol size.
For all the collision energies, we observe that the $M$ and $\sigma$
values increase,  whereas $S$ and $\kappa$ values decrease with
increasing \Npart. 
The dotted lines in the figure are central limit theorem (CLT) calculations
of the moments as a function of \Npart~\cite{CLT}, which assume 
independent emission sources.
These calculations follow the general trend of the data points.
However, deviations from the CLT have been observed for several data
points where the $\chi^2$ 
values are as large as 16.9 for 7 degrees of freedom.
This may imply correlated emission of particles.
The volume dependences of the moments are evident
from Fig.~\ref{fig1}, plotted as a function of \Npart, which are
cancelled in suitably constructed products of the moments.

\begin{figure}[tbp]
\centering
\includegraphics[width=0.49\textwidth]{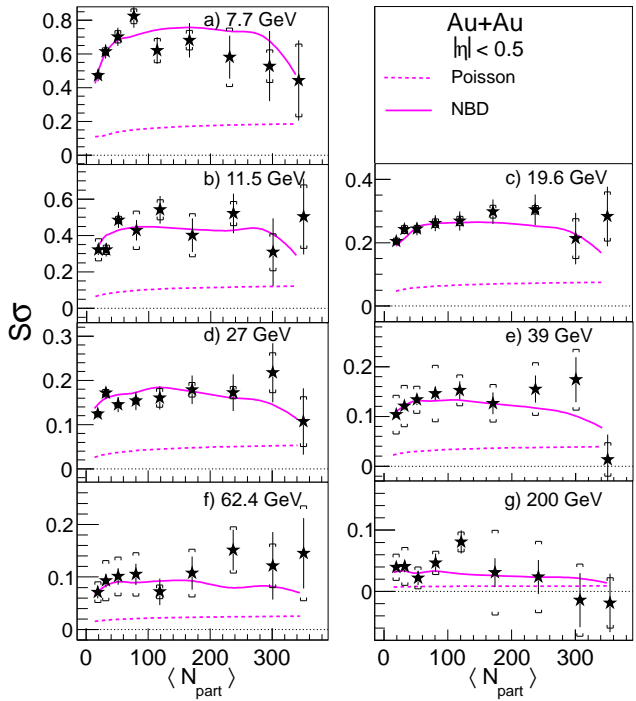}
\caption{(Color online) 
Centrality dependence of \Ss~in
Au+Au collisions at \sNN~=~7.7 to 200~GeV. The results are efficiency
and centrality bin width corrected.
Results from Poisson and the NBD baselines are superimposed.
The error bars are statistical and the caps represent systematic errors.}
\label{fig2}
\end{figure}

\begin{figure}[tbp]
\centering
\includegraphics[width=0.49\textwidth]{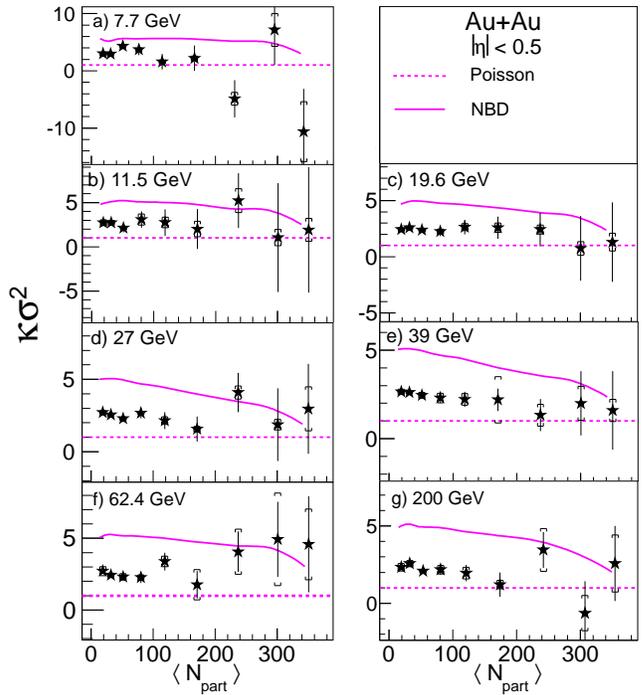}
\caption{(Color online) 
Centrality dependence of \KV~in
Au+Au collisions at \sNN~=~7.7 to 200~GeV. The results are efficiency
and centrality bin width corrected.
Results from Poisson and the NBD baselines are superimposed.
The error bars are statistical and the caps represent systematic errors.}
\label{fig3}
\end{figure}

In order to understand the nature of moments and their products, it is
essential to compare the experimental results with
baseline calculations. Two such calculations, one using the Poisson distribution
and the other the negative binomial distribution (NBD), have been
studied.  In case of the Poisson baseline, the
positive and negative charged particle multiplicities are randomly
sampled from their mean values, resulting in a Skellam net-charge
distribution~\cite{skellam}. 
The NBD baselines are constructed by using both the measured
mean values and variances of the positive and negative charged particles~\cite{NBD}.
Like the CLT, the Poisson and NBD baselines assume that the event by event
multiplicities of positive and negative particles are independent
random variables, i.e., completely uncorrelated. 
The Poisson and NBD assumptions result in 
different relationships between the moments of positive and negative  
particles.
These baselines may provide adequate references for the
moments of the net-charge distributions. Deviations from the baseline
values, if any, 
would help to observe possible non-monotonic behavior. 

Figures~\ref{fig2} and \ref{fig3} show the values of \Ss~and \KV,
respectively, plotted as functions of \Npart~for Au+Au at
seven collision energies. The data are corrected
for centrality bin width effect and detector efficiencies.
Results from Poisson and NBD 
baselines are superimposed in both figures.
The \Ss~values, shown in Fig.~\ref{fig2},
systematically decrease with increasing beam energy for all
centralities. The Poisson and NBD baselines are close to 
the data at \sNN~$=200$~GeV. The differences between the
baselines and the data increase with decreasing beam energy.
For low energies (\sNN~$\leq 27$~GeV),
the data  are systematically above the Poisson baselines 
by more than two standard deviations, whereas
the NBD baselines give a better description of the data.
Figure~\ref{fig3} shows that the values of \KV~at all 
energies and centralities are consistently
larger than the Poisson baselines and below the NBD baselines.
The NBD baselines are closer to the data than the Poisson, but fail to
quantitatively reproduce the experimental values. 
This is an indication of the existence of intra-event correlations of
positive and negative charged particles in the data, even within the
finite detector acceptance.

\begin{figure}[tbp]
\centering
\includegraphics[width=0.49\textwidth]{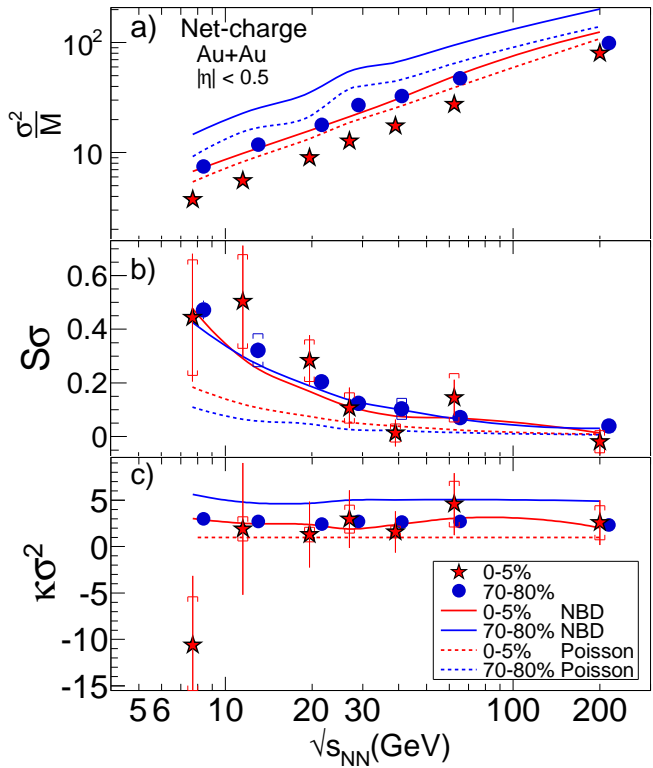} 
\caption{(Color online) Beam-energy dependence of (a) \MV,
(b) \Ss, and (c) \KV, after all corrections, for most central
(0-5\%) and peripheral (70-80\%) bins. 
The error bars are statistical and the caps represent systematic errors.
Results from the Poisson and the NBD baselines are superimposed.
The values of \KV~ for Poisson baseline are always unity.
}
\label{fig4}
\end{figure}

In Fig.~\ref{fig4}, we compare the beam-energy dependence of 
\MV, \Ss~and \KV~for two centrality
bins, one corresponding to most central (0-5\% bin) and the other 
to peripheral (70-80\% bin) collisions.
Results from the Poisson and NBD baselines are superimposed for both
of the centralities. All the results shown in this figure are 
efficiency and centrality bin width corrected.
The values of \MV~increase with increasing beam
energy, and are larger for peripheral collisions compared with 
the central collisions. In general, both the baseline calculations overestimate the data.
The \Ss~values are close to zero for \sNN~$= 200$~GeV, and
increase with decreasing beam energy for both centralities.
The Poisson baselines underestimate the \Ss~values in most of cases, whereas 
the NBD baselines are closer to the data. The peripheral data
are better described by the NBD baselines compared with the central
data points. 
The \KV~values for Poisson baselines are always unity. 
For peripheral collisions the \KV~values show almost no variation as a
function of beam energy and lie above the Poisson
baseline and below the NBD baseline. 
For central collisions,
within the statistical and systematic errors of the data, 
the \KV~values at all energies are
consistent with each other, except for \sNN~=~7.7~GeV.
The weighted mean of \KV~calculated for central collisions at all energies is $2.4\pm1.2$.
For central collisions, 
both of the baseline calculations follow the data points except for the
one at the lowest energy.
Deviations of the data points with respect to the baseline
calculations have been quantified in terms of the significance of
deviation, defined as, 
($|$Data--Baseline$|$)/($\sqrt{{\rm err}_{\rm stat}^2 + {\rm
   err}_{\rm sys}^2}$), where ${\rm err}_{\rm stat}$ and ${\rm err}_{\rm sys}$ are
the statistical and systematic errors, respectively.
These deviations remain within 2 in case of 
\Ss~and \KV~with respect to the corresponding Poisson
and NBD baselines. 
This implies that the products of moments do not 
show non-monotonic behavior as a function of beam energy.


The fluctuations of conserved quantities can be used to extract the
thermodynamic information on chemical freeze-out by comparing experimentally measured higher
moments with those from first-principle lattice QCD
calculations~\cite{freezeout_karsch}. 
Traditionally, by using the integrated
hadron yields, the first moment of the fluctuations, the chemical
freeze-out have been extracted from hadron resonance gas (HRG) models~\cite{Cleymans,andronic}.
However, higher-order correlation functions should allow stricter tests on the
thermal equilibrium in heavy-ion collisions.  
Calculations of freeze-out parameters based on preliminary experimental
data on moments of net-charge distributions have been obtained~\cite{swagato,freezeout_fodor}.
From the latest lattice~\cite{Borsyani} and HRG analyses~\cite{alba} using
the STAR net-charge and net-proton results for central Au+Au
collisions at 7.7 to 200 GeV, the extracted freeze-out temperatures
range from 135 to 151 MeV and \muB~values range from 326 to 23 MeV. 
The errors in these calculations increase from 2\% to 10\% 
as a function of decreasing beam energy, which is mostly due 
to the statistical uncertainty in the experimental measurements.
More details can be found in \cite{Borsyani,alba}.
Note that this is the first time that the experimentally measured
higher moments are used to 
determine the chemical freeze-out conditions in high-energy nuclear
collisions.  The freeze-out temperatures obtained from the higher
moments analysis are 
lower with respect to the traditional method~\cite{Cleymans,lokesh}.  
This difference could indicate a higher sensitivity to freeze-out in
the higher moments, which warrants further investigation.

In summary, the first results of the moments of
net-charge multiplicity distributions for $|\eta| < 0.5$ as a function of
centrality for Au+Au collisions at seven collision energies from
\sNN~=~7.7 to 200~GeV are presented.
These data can be used to explore the nature of the QCD phase
transition and to locate the QCD critical point. 
We observe that the \MV~values increase monotonically with increasing beam energy.
Weak centrality dependence is observed for both \Ss~and \KV~at all
energies. The \Ss~values increase with decreasing beam energy,
whereas \KV~values are uniform except at the lowest beam energy.
Most of the data points show deviations from Poisson baselines. 
The NBD baselines are closer to the data than Poisson, but do not
quantitatively reproduce the data, implying the importance of
intra-event correlations of the multiplicities 
of positive and negative particles in the data.
Within the present uncertainties, no 
non-monotonic behavior has been
observed in the products of moments as a function of collision energy.
The measured moments of net-charge multiplicity distributions provide unique information about the freeze-out parameters by directly comparing with theoretical model calculations.
Future measurements with high statistics data will be needed for precise 
determination of freeze-out conditions and to
make definitive conclusions regarding the critical point. 

\medskip

We thank M. Asakawa, R. Gavai, S. Gupta, F. Karsch, V. Koch, S. Mukherjee,
K. Rajagopal, K. Redlich and M. A. Stephanov for discussions related to this work.
We thank the RHIC Operations Group and RCF at BNL, the NERSC Center at
LBNL, the KISTI Center in Korea, and the Open Science Grid consortium
for providing resources and support. This work was supported in part
by the Offices of NP and HEP within the U.S. DOE Office of Science,
the U.S. NSF, CNRS/IN2P3, FAPESP CNPq of Brazil,  the Ministry of
Education and Science of the Russian Federation, NNSFC, CAS, MoST and
MoE of China, the Korean Research Foundation, GA and MSMT of the Czech
Republic, FIAS of Germany, DAE, DST, and CSIR of India, the National
Science Centre of Poland, National Research Foundation
(NRF-2012004024), the Ministry of Science, Education and Sports of the
Republic of Croatia, and RosAtom of Russia.



\begin{thebibliography}{00}
\bibitem{CLT} According to the central limit theorem, 
$M$, $\sigma^{2}$, $S$ and $\kappa$ are proportional to
$\langle N_{{\rm part}}  \rangle$, $\sqrt{\langle N_{{\rm part}}  \rangle}$, 
$1/\sqrt{\langle N_{{\rm part}}  \rangle}$, $1/{\langle N_{{\rm part}}  \rangle}$, respectively.
\bibitem{skellam}
The difference of two independent Poisson distributions is a Skellam
distribution, for which  $\frac{\sigma^{2}}{M} = \frac
{\langle N_{+} \rangle +  \langle N_{-} \rangle}
{\langle N_{+} \rangle -  \langle N_{-} \rangle}$,
$S\sigma = \frac
{\langle N_{+} \rangle -  \langle N_{-} \rangle}
{\langle N_{+} \rangle +  \langle N_{-} \rangle}$, and
$\kappa\sigma^{2} = 1$.
\end{thebibliography}
\end{document}